\documentclass[conference]{IEEEtran}

\usepackage{cite}

\ifCLASSINFOpdf
  \usepackage[pdftex]{graphicx}
\else
  \usepackage[dvips]{graphicx}
\fi
\usepackage[cmex10]{amsmath}
\usepackage{amssymb}  
\usepackage{amsthm}

\newtheorem{defn}{Definition}
\newtheorem{cor}{Corollary}
\newtheorem{lem}{Lemma}

\usepackage{mdwlist}

\begin{document}
%
\title{Feasibility of Simultaneous Information and Energy Transfer in LTE-A
Small Cell Networks}

\author{\IEEEauthorblockN{Hongxing Xia}
\IEEEauthorblockA{School of Computer and Information Science\\
Hubei Engineering University\\
Xiaogan, P.R.China\\
Email: xiahx@hbeu.edu.cn}
\and
\IEEEauthorblockN{Balasubramaniam Natarajan\\ and Chang Liu}
\IEEEauthorblockA{Department of Electrical and Computer Engineering\\
Kansas State University, Manhattan\\
Email: bala@k-state.edu}

}


%


\maketitle

\begin{abstract}
Simultaneous information and energy transfer is attracting much attention
as an effective method to provide green energy supply for mobiles.
However the very low power level of the harvested energy from RF spectrum
limits the application of such technique. Thanks to the improvement
of sensitivity and efficiency of RF energy harvesting circuit as well
as the dense deployment of small cells base stations, the SIET becomes
more practical. In this paper, we propose a unified receiver model
for SIET in LTE-A small cell base staion networks, formulate the feasibility
problem with Poisson point process model and analysis the feasibility
for a special and practical senario. The results shows that it is
feasible for mobiles to charge the secondary battery wih harvested
energy from BSs, but it is still infeasible to directly charge the primary
battery or operate without any battery at all.\end{abstract}

\section{Introduction}

There are two major roles for RF energy. The most
important use among them is in providing telecommunications services
to the public, industry and government. The non-communications uses
of RF energy mainly include heating ,radar and wirelss power transfer
(WPT). Due to the shotage of fossil fuels and the crisis of environment,
WPT and energy harvesting have received considerable attention as
methods of addressing environmental problems \cite{SUD11,VAR08}.
There are two ways for transmitting information and wireless power:
\emph{single tone }and \emph{multi-tone} methods \cite{KAW13}. The
former uses only one carrier to transmit information and power simultaneously;
The latter method transmits the information and energy seperatly with
two distinct carrier frequencies. Since the spectrum resources is
also very limited today, people spend more energy on the research
of simultaneous information and energy transfer (SIET) \cite{LIU13,SIX13}.

As harvesing energy from ambient RF signal is free and unlimited,
the SIET has recently drawn a great attention. A point-to-point transfer
with signle antenna is studied in \cite{LIU13}, their work investgates
when the receiver should switch between the two modes of information
decoding and energy harvesting based on the instaneous channel and
interference condition. In \cite{ZHANG13}, a simultaneous wireless
information and power transfer with MIMO broadcast system is considered.
To optimal the the transfer strategy to achieve tradeoffs for maximal
information rates versus energy transfer, the boundary of rate-energy
region is chracterized. They also propose two practical designs for
co-located receiver called\emph{ time switching} and \emph{power splitting}.
Howerver, from the perspective of practicalbility, the key problem
for SIET is whether the energy is strong enough to sustain the mobiles.
From Fig 4 in \cite{ZHANG13}, it can be found that the maximal harvested
energy will not exceed $0.6mW$ for a $4\times4$ MIMO broadcast system,
even though the information rate is lowered to $0$. The work of \cite{huang12}
propose a more practical design for cellular networks to tansfer wireless
power: delpoying a new type base stations called power beacons (PBs) to
deliver energy to mobile devices by microwave radiation. The PBs are
deployed as a homogeneous Poisson point process (PPP) with a certain
density. It is proved in this work that the density and transmit power
of the PBs must satisfy some condition to meet the outage constraint
of the mobiles. However, this scheme needs extra construction of PBs
except for common base stations and it is economically infeasible.

Thanks to the improvement of sensitivity and efficiency of RF energy
harvesting circuit \cite{Dol10}, the simultaneous information and
energy transfer (SIET) is becoming more and more practical. More importantly,
with densely deployed small cells base stations (recently proposed
to LTE-A) \cite{Naka13}, the closer distance to the radio emitter
can greatly improve the energy transfer efficiency. Besides, the interference
from other BSs can also contribute to the energy harvesting, which
means all signals in the air is useful. Encouraged by above observation,
we propose a practical receiver model for SIET in a homogeneous small
cell networks. Based on this model we focus on the feasibility study
of the SIET using stochastic geometry method. The main contributions
of our work are as follows:
\begin{itemize*}
\item Propose a unified receiver model for mobiles of PPP BS-deployed LTE-A
small cell networks which can decode information and harvest energy
simultaneously. Such model considers the user activity level as well
as a flexible power allocation factor for energy-harvesting and information-decoding.
\item Formulate the distribution of the energy harvested from a PPP deployed
base stations for the first time. Also the definition of \emph{efficient
energy harvesting (EEH) probability} is first proposed .
\item Formulate the feasibility of SIET in small cell networks as maximization
of the EEH probability conditioned on constraints of coverage probability
and density and power limitation of BSs.
\item The feasibility problem for a special case which has limited interference
and path loss exponent as $4$ is analytically solved. The result
shows that it is still infeasible for the harvested energy to compensate
the basic energy consumption of a low-power mobile, but it is feasible
to charge the secondary battery of a hybrid-battery supplied terminal.
\end{itemize*}
In section \ref{sec:System-Model}, the system model is proposed.
Section \ref{sec:Preliminary} provide the preliminary of coverage
and effcient energy harvesting. The feasibility problem is formulated
and solved in section\ref{sec:Problem-Formulation}, and considering
real senario the feasbility is analysed in section \ref{sec:Feasibility-Analysis}.
The conclusion and future research direction is given in section \ref{sec:Conclusion}.

\section{System Model\label{sec:System-Model}}

We consider a homogeneous small cells networks with base stations
arrranged according to Possion point process (PPP) $\Phi$ of intensity
$\lambda$ . We also assume the mobile users loactaed accroding to
some independent stationary point process and each mobile user is
associated with the closest base station notated as $b_{c}$.

For simlicity and tractability, we set the base station and associated
mobile user experience fading channel with mean $1$. The standard
power loss propagation model is used with path loss exponent $\alpha>2$
. Besides, all the base staions transmit power is set to be $P$.
Then the received power at a typical mobile user with a distance $r$
from its corresponding base station is $hr^{-\alpha}$ where the random
variable $h$ follows exponential distribution with mean $P$, i.e.
$h\sim\mathrm{exp}(1/P)$ .

\begin{figure}[h]
\centering
\includegraphics[scale=0.6]{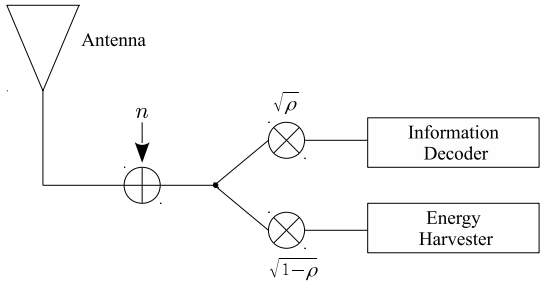}\caption{Receiver Model of SIET system\label{fig:Receiver}}
\end{figure}

We employ a receiver model as figure \ref{fig:Receiver} . The received
raw power is splitted into two streams, one stream is fed into the
information decoder while the other into the energy harvestor. The
power splitting facor is denoted as $\rho$. The white Gaussian noise
introbuced by the receiveing antenna is represented as $n$,where
$n\sim\mathcal{CN}(0,\sigma^{2})$.

We focus on the downlink SIET in this paper.Obvisoualy the receiver
have two states in the downlink diretion. One state is that the user
is scheduled by the BS and have to decode information and harvest
energy simultaneously. Under the other state, the user is inactive
and solely harvest energy from the ambient RF. To maximize the utilization
of RF energy, we suppose the power splitting factor can be adjusted
in light of the user's state. We model the user activity as a two
state markov process and assume the probability of user being active
and idle is $\epsilon$ and $1-\epsilon$, respectively. Thus the
adpative power splitting factor can be described as $\rho\centerdot\mathbf{1}(\mathrm{user\,\, is\,\, active})$.
It means that the user feed all the received energy to the energy
harvester when the user is inactive, as in such case there is no information
need to be decoded.

Then the received power $P_{0}$ at the origin before splitting can
be represented as:

\begin{equation}
P_{0}=\begin{cases}
hr^{-\alpha}+\sum_{i\in\Phi/b_{c}}hR_{i}^{-\alpha}+\sigma^{2} & \mathrm{user\,\, is\,\, active}\\
\sum_{i\in\Phi}hR_{i}^{-\alpha}+\sigma^{2} & \mathrm{user\,\, is\,\, idle}
\end{cases}
\end{equation}

Where $R_{i}$ denotes the distances from the $i$'s base station
to a typical user located at the origin.

\section{Preliminary\label{sec:Preliminary}}

\subsection{Coverage}
\begin{defn}
\emph{Coverage: A user is in coverage when its SINR from its nearest
BS is larger than some threshold T and it is dropped from the network
for SINR below T.}
\end{defn}
According to definition 1, the coverage probability of homogeneous
network can be formulated as

\begin{equation}
p_{c}(T,\lambda,P,\alpha,\rho)\triangleq\mathbb{P}[\textrm{SINR}>T]\label{eq:pc}
\end{equation}

The SINR of the mobile user at a random distance $r$ from its associated
base station can be expressed as:

\begin{equation}
\textrm{SINR}=\frac{\rho hr^{-\alpha}}{\sigma^{2}+\rho\sum_{i\in\Phi/b_{c}}hR_{i}^{-\alpha}}\label{eq:sinr}
\end{equation}

\subsubsection{Distance to Nearest base Sattion}

The probability density function(pdf) of $r$ can be derived using
the fact that the null probability of a 2-D Possion process in an
area $A$ is $\exp(-\lambda A)$

\[
\mathbb{P}[r>R]=\mathbb{P}[\textrm{No BS closer than R}]=e^{-\lambda\pi R^{2}}
\]
Then the pdf of $r$ is found as:

\begin{equation}
f_{r}(r)=e^{-\lambda\pi r^{2}}2\pi\lambda r
\end{equation}

\subsubsection{Average Coverage Probability}

In order to calculate the coverage probability, we first restate a
known result from the stachastic geometry theory\cite{Muk11}. Then
this result are employed to derive the complementary cumulative distribution
(ccdf) of SINR for a typical user.
\begin{cor}
For a homogeneous cellular networks of which the BS's positions follow
PPP with indensity $\lambda$, the interference at the origin from
those base stations at least $r$ away from the user can be formulated
as:
\begin{equation}
I(r)=\sum_{i:R_{i}>r}hR_{i}^{-\alpha},\label{eq:Ir}
\end{equation}
 where $h$ follows exponential distribution with parameter $\mu$
and indepentent of the distance $\{R_{i}\}$. Then the Laplace Transform
of $I(r)$ at any $s>0$ is
\begin{equation}
\mathcal{L}_{I(r)}(s)=\exp[-\pi\lambda(s/\mu)^{2/\alpha}G(r^{2}(s/\mu)^{-2/\alpha})],\label{eq:LIS}
\end{equation}

where
\begin{equation}
G(y)=\int_{y}^{\infty}\frac{dx}{1+x^{\frac{\alpha}{2}}}=\begin{cases}
\pi/2-\arctan y, & \alpha=4,\\
_{2}F_{1}(1,\frac{2}{\alpha};1+\frac{2}{\alpha};-x^{\frac{\alpha}{2}})x|_{y}^{\infty}, & \alpha\neq4,
\end{cases}
\end{equation}

and $_{2}F_{1}(a,b;c;z)$is the hypergeometric function.

For special case with $r=0$, $\mathcal{L}_{I(0)}(s)=\exp[-\frac{2\pi^{2}\lambda}{\alpha}\left(\frac{s}{\mu}\right)^{\frac{2}{\alpha}}\csc\left(\frac{2\pi}{\alpha}\right)]$\end{cor}
\begin{IEEEproof}
The corollary is straight derived from Corllary 1 in \cite{Muk11}
by subsititue $X_{i}$ with $h$ and $\mu$ with $1/\mu$ for consistency
with our notation custom.\end{IEEEproof}
\begin{lem}
To exmaine the overall coverage performance of the network, the average
coverage probability over the plane can be presented as:

\begin{equation}
\mathcal{P}_{c}(T,\lambda,P,\alpha,\rho)=2\pi\lambda\int_{r>0}e^{-\pi\lambda r^{2}-Tr^{\alpha}\sigma^{2}/\rho P}\mathcal{L}_{I(r)}(Tr^{\alpha}/P)rdr
\end{equation}

where $\mathcal{L}_{I(r)}(s)$ is the Laplace transform of random
variable $I(r)$ evaluated at $s$ conditioned on the distance to
the closest BS from the origin.\end{lem}
\begin{IEEEproof}
Subsititute \ref{eq:sinr} into \ref{eq:pc-1} follows:

\begin{eqnarray*}
p_{c}(T,\lambda,P,\alpha,\rho) & = & \mathbb{P}[\frac{\rho hr^{-\alpha}}{\sigma^{2}+\rho I(r)}>T|r]\\
 & = & \mathbb{E}_{I(r)}[\mathbb{P}[h>T\rho^{-1}r^{\alpha}(\sigma^{2}+\rho I(r))|r,I_{r}]\\
 & \overset{(a)}{=} & \mathbb{E}_{I(r)}[\exp(-T(\rho P)^{-1}r^{\alpha}(\sigma^{2}+\rho I(r))|r]\\
 & = & e^{-Tr^{\alpha}\sigma^{2}/\rho P}\mathcal{L}_{I(r)}(Tr^{\alpha}/P),
\end{eqnarray*}

where (a) follows that $h\sim\exp(1/P)$. The average covergae probability
over the plane can be expressed as

\begin{eqnarray}
\mathcal{P}_{c}(T,\lambda,P,\alpha,\rho) & = & \int_{r>0}p_{c}(T,\lambda,\alpha)f_{r}(r)dr\nonumber \\
 & = & 2\pi\lambda\int_{r>0}e^{-\pi\lambda r^{2}-\frac{Tr^{\alpha}\sigma^{2}}{\rho P}}\mathcal{L}_{I(r)}(\frac{Tr^{\alpha}}{P})rdr.\label{eq:pc-1}
\end{eqnarray}

Then we obtain the result.
\end{IEEEproof}

\subsection{Efficient Energy Harvesting}
\begin{defn}
\emph{Efficient Energy Harvesting (EEH): A user is able to harvest
usable energy from ambient RF only if its received energy is larger
than certain threshold $\Theta$ due to the constraint of energy harvesting
circuit.}
\end{defn}
Then the EEH probability $p_{\mathrm{\mathsf{\mathrm{eeh}}}}(\Theta,\lambda,\alpha,\rho)$
of a typical user located at the origin can de defined as:

\begin{equation}
p_{\mathit{\mathrm{eeh}}}(\Theta,\lambda,P,\alpha,\rho)\triangleq\mathbb{P}[E_{h}>\Theta],
\end{equation}

Averaging the EEH probability over distance as well as the user state
can derive

\begin{equation}
\mathcal{P}_{\mathrm{eeh}}(\Theta,\lambda,P,\alpha,\rho)\triangleq\mathbb{E}_{r,us}[\mathbb{P}[E_{h}>\Theta|r,us]].\label{eq:eeh}
\end{equation}

\begin{lem}
\label{lem:peeh}The average probability of efficient energy harvesting
of a typical randomly located user in the small cell networks is

\begin{equation}
\mathrm{\mathcal{P}_{eeh}}(\Theta,\lambda,P,\alpha,\rho)=1-\epsilon F_{I(0)}(\Theta-\sigma^{2})-(1-\epsilon)F_{I(0)}(\Theta-\sigma^{2}),\label{eq:paeeh}
\end{equation}

where $F_{I(0)}(x)=\mathcal{L}_{s}^{-1}\left\{ \frac{1}{s}\exp[-\frac{2\pi^{2}\lambda}{\alpha}\left(\frac{s}{\mu}\right)^{\frac{2}{\alpha}}\csc\left(\frac{2\pi}{\alpha}\right)]\right\} (x)$.\end{lem}
\begin{IEEEproof}
For energy harvesting, there is no differnece between the cases with
active state and idle state. Since the energy harvestor does not need
to extract information from it's corresponding BS, we then can treat
the harvested energy on both cases as interference from all the base
sattions. According to the definition of (\ref{eq:Ir}), the harvested
energy before power splitting can be expressed as $I(0)+\sigma^{2}$
and does not depend on the distance $r$. Note that the distance $r$
only affects which base station should be connected but not the whole
interfernce in the plane. Now (\ref{eq:eeh}) can be rewritten as

\begin{equation}
\begin{aligned} & \mathcal{P}_{\mathrm{eeh}}(\Theta,\lambda,P,\alpha,\rho)\\
= & \mathbb{E}_{us}[\mathbb{P}[E_{h}>\Theta]]\\
= & \epsilon\mathbb{P}[(I(0)+\sigma^{2})(1-\rho)>\Theta]+(1-\epsilon)\mathbb{P}[I(0)+\sigma^{2}>\Theta]\\
= & \epsilon\mathbb{P}[I(0)>\frac{\Theta-\sigma^{2}}{1-\rho}]+(1-\epsilon)\mathbb{P}[I(0)>\Theta-\sigma^{2}]\\
= & \epsilon(1-F_{I(0)}(\frac{\Theta-\sigma^{2}}{1-\rho}))+(1-\epsilon)(1-F_{I(0)}(\Theta-\sigma^{2}))\\
= & 1-\epsilon F_{I(0)}(\frac{\Theta-\sigma^{2}}{1-\rho})-(1-\epsilon)F_{I(0)}(\Theta-\sigma^{2})
\end{aligned}
,\label{eq:peehb}
\end{equation}

where $F_{I(0)}(x)$ is the cdf of $I(0)$. There is no close-form
expression for the cdf (pp97,\cite{martin12}), but we can recover
the cdf by inversing the Laplace transform of $I(0)$ as following:
\begin{eqnarray*}
F_{I(0)}(x) & = & \mathcal{L}_{s}^{-1}\left\{ \frac{\mathcal{L}_{I(0)}(s)}{s}\right\} (x)\\
 & = & \mathcal{L}_{s}^{-1}\left\{ \frac{1}{s}\exp[-\frac{2\pi^{2}\lambda}{\alpha}\left(\frac{s}{\mu}\right)^{\frac{2}{\alpha}}\csc\left(\frac{2\pi}{\alpha}\right)]\right\} (x).
\end{eqnarray*}

Then Lemma \ref{lem:peeh} is proved.
\end{IEEEproof}

\section{Problem Formulation\label{sec:Problem-Formulation}}

In this paper, we mainly study the feasibility of the SIET in small
cell networks. Intutively, the denser the BSs are deployed, the more
power the user can harvest and then the feasibillity of SIET is increased.
However for communication purpose, excessive denser BSs will not bring
better link quality but more severe interference. So we consider an
density-limited small cell networks with ability of concurent transmission
of energy and information. That is to maximize the Efficient Energy
Harvesting probability under constraint of coverage probability ,
BS's transmit power and BS-deployment density . The problem is formulated
as below:

\begin{eqnarray}
\mathrm{\mathbf{P1}:\underset{\mathit{p,\lambda}}{\mathbf{max}}} &  & \mathcal{P}_{eeh}(\Theta,\lambda,P,\alpha,\rho)\label{eq:P1}\\
\mathrm{\mathbf{s.t.}} &  & \mathcal{P}_{c}(T,\lambda,P,\alpha,\rho)\geqslant\mu\label{eq:p1c1}\\
 &  & P\leqslant P_{\mathrm{max}}\label{eq:p1c2}\\
 &  & \lambda\leqslant\lambda_{\mathrm{max}},\label{eq:p1c3}
\end{eqnarray}

for given thresholds of SINR($T$) and energy harvesting threshold
($\Theta$), where $\mu$ is the minimum coverage probability, $P_{max}$
and $\lambda_{max}$ is the maximum transmit power of the small cell
BSs and the maximum BS-deployment density of the networks, respectively.
The setting of EEH threshold $\Theta$ and SNR threshold is based
on different service quality requirement. Unfortunately, the problem
(\ref{eq:P1}) is intractable due to integration form of $\mathcal{P}_{c}(T,\lambda,P,\alpha,\rho)$
and inverse Laplace transform expression of $\mathcal{P}_{eeh}(\Theta,\lambda,P,\alpha,\rho)$
. In the remaining part of this paper, we simplify the problem to
a special case with $\alpha=4$ and $\sigma^{2}=0$, where it leads
to a closed-form expressions for $\mathcal{P}_{\mathrm{c}}$ and $\mathcal{P}_{\mathrm{eeh}}$
. Note that we intend to gain an insight into the feasibility of SIET
in small cell networks, thereby this simplification does not weaken
the focus of this paper.

\subsection{Interference Limit Case with $\alpha=4$}

When we set $\alpha=4$ and $\sigma^{2}=0$, the Laplace transform
of $I(r)$ in (\ref{eq:LIS}) is simplified to
\begin{equation}
\mathcal{L}_{I(r)}(s)=e^{-\pi\lambda\sqrt{sP}(\frac{\pi}{2}-\arctan\frac{r^{2}}{\sqrt{sP}})}.\label{eq:LIS-Simplified}
\end{equation}

Introducing (\ref{eq:LIS-Simplified}) into (\ref{eq:pc-1}) can get

\begin{equation}
\mathcal{P}_{c}(T)=\frac{1}{1+\sqrt{T}(\frac{\pi}{2}-\arctan\frac{1}{\sqrt{T}})},\label{eq:pc-simplified}
\end{equation}

where the coverage probability does not depend on $\lambda$ or $\rho$.
This means that the constraint (\ref{eq:p1c1}) can be removed in
such case.

Next we introduce these special $\alpha$ and $\sigma^{2}$ into (\ref{eq:paeeh})
and simplify the average effective energy harvesting probability $\bar{p}_{\mathrm{eeh}}$
to

\begin{equation}
\mathcal{P}_{\mathrm{eeh}}(\Theta,\lambda,P,\rho)=\epsilon\,\mathrm{erf}(\frac{\pi^{2}\lambda}{4}\sqrt{\frac{P(1-\rho)}{\Theta}})+(1-\epsilon)\mathrm{erf}(\frac{\pi^{2}\lambda}{4}\sqrt{\frac{P}{\Theta}}),\label{eq:peeh-simplified}
\end{equation}

where $\mathrm{erf}(x)=2/\sqrt{\pi}\int_{0}^{x}e^{-t^{2}}dt$ is the
standard error function.
\begin{IEEEproof}
Omitted due to page limit.
\end{IEEEproof}

\subsection{Solution in Secial Case}

Using the simpified expression of EEH probability $\bar{p}_{eeh}$,
and removing the constraint (\ref{eq:p1c1}) the problem (\ref{eq:P1})
degrades to

\begin{eqnarray}
\mathrm{\mathbf{P2}:\underset{\mathit{p,\lambda}}{\mathbf{max}}} &  & \mathcal{P}_{\mathrm{eeh}}(\Theta,\lambda,P,\rho)\label{eq:P2}\\
\mathrm{\mathbf{s.t.}} &  & P\leqslant P_{\mathrm{max}}\\
 &  & \lambda\leqslant\lambda_{\mathrm{max}}.
\end{eqnarray}

\begin{figure}[tbh]
\begin{centering}
\includegraphics[scale=0.48]{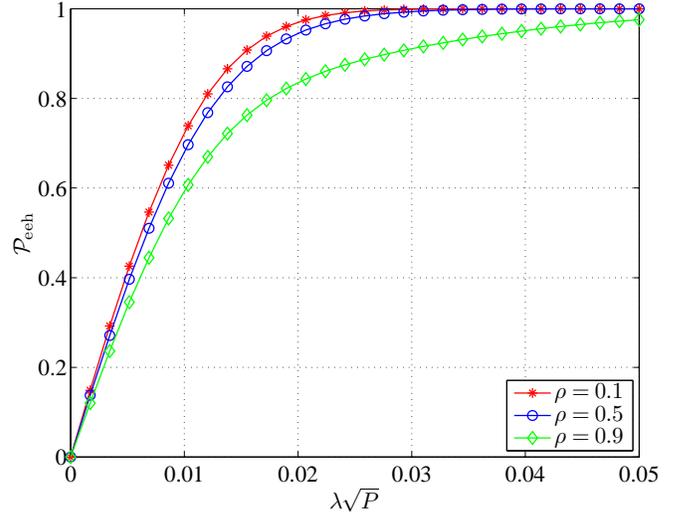}
\par\end{centering}

\caption{Curve of $\mathcal{P}_{\mathrm{eeh}}$ over $\lambda\sqrt{P}$ with
different power splitting factors,$\epsilon=0.3$ and $\Theta=1\mathrm{mw}$.\label{fig:Curve-of-Peeh}}

\end{figure}

By carefully looking at (\ref{eq:peeh-simplified}) we can find that
given the EEH threshold $\Theta$ , the energy splitting factor $\rho$
and the user active probability $\epsilon$, the Efficient Energy
Harvesting probability is monotonously increased with $\lambda\sqrt{P}$.
This implies that from the perspective of harvesting energy, quadratic
increasing of transmit power is equivalent to liniear increasing of
network density, which coincides with the result of interference analysis
in \cite{martin12}. The curve of $\bar{\mathcal{P}}_{\mathrm{eeh}}$
with regard to $\lambda\sqrt{P}$ is dipcted in figure (\ref{fig:Curve-of-Peeh}),
assuming $\epsilon=0.3$ and $\Theta=1\mathrm{mw}$. With above observation,
the solution of (\ref{eq:P2}) is straightforward and the optimal
value of EEH probability is achieved when $P$ and $\lambda$ take
their maximum values synchronously. Due to the equivalence of effects
of $\lambda$ and $\sqrt{P}$ on energy-harvesting, we can set transmit
power $P$ as a typical constant value and study the maximum EEH probability
with distinct BS-deployment density. For most small cell base stations,
the transmit power would not exceed $1W$, then it is reasonble to
set $P=1W$. And the BSs density $\lambda$ under such assumption
is defined as \textit{standard base station density}.

\begin{defn}
\emph{\label{Standard-base-station}Standard base station density
($\lambda_{s}$): For a homogeneous PPP cellular network, if the transmit
power of all the base stations is $1$W, then the density of the PPP
cellular networks is called standard base station density.}
\end{defn}
It is worth noting that standard base station density is defined for
easing the analysis of energy-harvesting and interference. Since the
equivalence among the variation of density and transmit power, the
result with standard density can be readily extended to non-unit-transmit-power
case. With definition \ref{Standard-base-station}, the problem (\ref{eq:P2})
can be further simplified as follows:

\begin{eqnarray}
\mathrm{\mathbf{P3}}:\underset{\mathit{\lambda_{s}}}{\mathbf{max}} &  & \mathcal{P}_{\mathrm{eeh}}(\Theta,\lambda_{s})\label{eq:p3}\\
\mathrm{\mathbf{s.t.}} &  & \lambda_{s}\leqslant\lambda_{\mathrm{max}}.\nonumber
\end{eqnarray}

where
\begin{equation}
\mathcal{P}_{\mathrm{eeh}}(\Theta,\lambda_{s})=\epsilon\,\mathrm{erf}(\frac{\pi^{2}\lambda_{s}}{4}\sqrt{\frac{1-\rho}{\Theta}})+(1-\epsilon)\mathrm{erf}(\frac{\pi^{2}\lambda_{s}}{4\sqrt{\Theta}}).\label{eq:peehlast}
\end{equation}
 The objection function is obviously increasing with larger $\lambda_{s}$,
then the EEH probability can get maximal value as $\mathcal{P}_{\mathrm{eeh}}^{*}(\Theta,\lambda_{\mathrm{max}})$.

\section{Feasibility Analysis\label{sec:Feasibility-Analysis}}

For analysising the feasibility of simultaneous information and energy
transforming, the most important parameter is $\Theta$, or the threshold
that the received power could be useful for sustaining the circuit
of the wireless terminal. This threshold is closely related to the
converter efficiency of the RF energy harvester and the lowest needed
power for maintaining the operation of a terminal. From the perspective
of energy supply, there are different levels of requirement for the
harvested ernergy.
\begin{enumerate*}
\item \emph{Charging the secondary battery}: the harvested energy can charge
the built-in battery and prolong the stand-by time of the device.
For example, the wireless device can have a hybrid-battery power supply
system, the primary battery is charged by the grid and the secondary
battery is charged by the RF energy harvester. In such case, the needed
power from energy harvester can be just less than the maintenance
power of the device.
\item \emph{Sustaining the basic system}: the harvested energy can completely
compensate the power consumption of the device when it has not communication
or other computation task. To deal with such computation tasks, it
is neccessary to build a grid-charged battery in the system. Compared
with level 1, the improvement is less battery, which means samller
volume and lighter weight of the device, and longer lasting power.
\item \emph{Battery-free:} if the harvested energy is large enough, the
device can be battery-free and the energy needed to support all the
tasks of the device entirely comes from the ambient spectrum.
\end{enumerate*}
As the power consumption of distinct devices is greatly varied, it
is impossible to find a unified standard for all kinds of terminals.
If we assume the maintenance power is $p_{\mathrm{m}}$, the\emph{
availability factor} is $\zeta$, then we can use $\zeta p_{\mathrm{m}}$
to describe the needed power for above three levels. More specifically,
power level of charging the secondary battery can be represented by
$\zeta p_{\mathrm{m}}$ with $0<\zeta<1$, sustaining the basic system
with $\zeta=1$ and battery-free with $\zeta\gg1$. Larger $\zeta$
implies more availability of the RF energy from the small cell base
stations. Integrating above analysis the threshold of harvested energy
before converter can be calculated as:

\begin{equation}
\Theta=\frac{\zeta p_{\mathrm{m}}}{\eta},
\end{equation}

where $\eta$ is the converter efficiency. According the recent development
of the RF energy harvester \cite{Scor13}, the peak efficiency can
achieve 60\% and average efficiency 40\% in the 840-975 MHz band.
As the converter efficiency for higher frequency (e.g. the frequency
over which the cellular communication operates) is not clear till
now, we will study the effects of different $\eta$ on the feasibility
of SIET. For the other key parameter $p_{\mathrm{m}}$, experimental
measurement shows that the typical maintenance power for a smart phone
is as much as $0.02W$ (3G) or $0.03W$ (GSM)\cite{Balasub09}. The
maintenance power of the LTE-A, which is concerned in this paper,
is believed not to exceed $0.02W$. In view of this observation, we
set $p_{\mathrm{m}}=0.02W$ for the mobile terminal of small cell
networks in the remaining part. It is noteworthy that the maintenance
power is greatly dependent on the hardware and operating system of
the mobile, e.g. an smartphone with ARM920T CPU and Android 1.5 operating
system will cost $0.068W$ for sustaining the basic system \cite{Carroll10}.
However, we only concern the feasibility of SIET in cellular communication
and thus the lowerest maintenance power is considered in this work.
Next we discuss two types of small cell networks according to different
density of the distribution of BSs: the small cell networks with $\lambda_{\mathrm{max}}=10^{-4}$
and the dense small cell networks with $\lambda_{\mathrm{max}}=10^{-2}$.
We also studied the relationship between the maximum BS-deployment
density and availability factor conditioned on constant average EEH
probability.

\subsection{Avarage EEH probability - Availability Factor Region}

\begin{figure}[tbh]
\begin{centering}
\includegraphics[width=21pc]{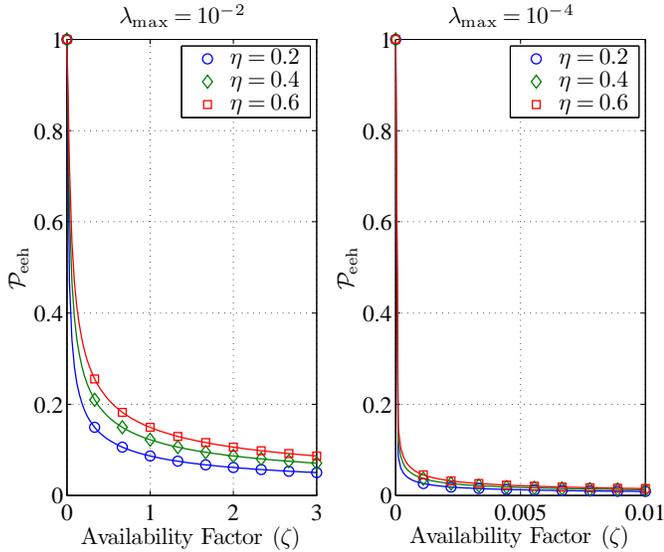}
\par\end{centering}

\caption{Maximal average EEH probability over availability factor, which is
computed using (\ref{eq:peehlast}) with $\epsilon=0.3$, $\rho=0.1$
and $p_{m}=0.02W$\label{fig:Maximal-average-EEH}}
\end{figure}

To study the feasibility of SIET on different BSs density, we dipict
the average EEH probability and availability factor region as figure
\ref{fig:Maximal-average-EEH}. From the figure it can be found that
when $\lambda_{{\rm max}}=10^{-2}$, which represents a type of dense
small cell networks, the average EEH probability only reaches $0.2$
for feasible availability factor, even with the converter efficiency
as much as $0.6$. For a practical application scenario, the avarage
EEH probability $\mathcal{P}{\rm _{{\rm eeh}}}$ should be at least
larger than 0.5 where the corresponding availability factor is $\zeta\in(0,1)$.
That means even with a very dense BS-deployment, the harvestd energy
from the BSs can only charge the secondary battery for s hybrid-battery
powered device. And the charging efficiency is proportional to the
availability factor.

For a more practical senario with small cell BS density $\lambda_{{\rm max}}=10^{-4}$
, the availability will not be larger than 0.01 even though
the avarage EEH probability is far less than 0.2, as shown in the
right part of figure \ref{fig:Maximal-average-EEH} . So in this case
harvesting energy from small cell BSs is impossible under current
converter efficiency of the havresters and power consumption of cell
phones.

\subsection{Maximal BS-deployment density over availability under constant $\mathcal{P}_{{\rm eeh}}$}

\begin{figure}[tbh]
\begin{centering}
\includegraphics[width=21pc]{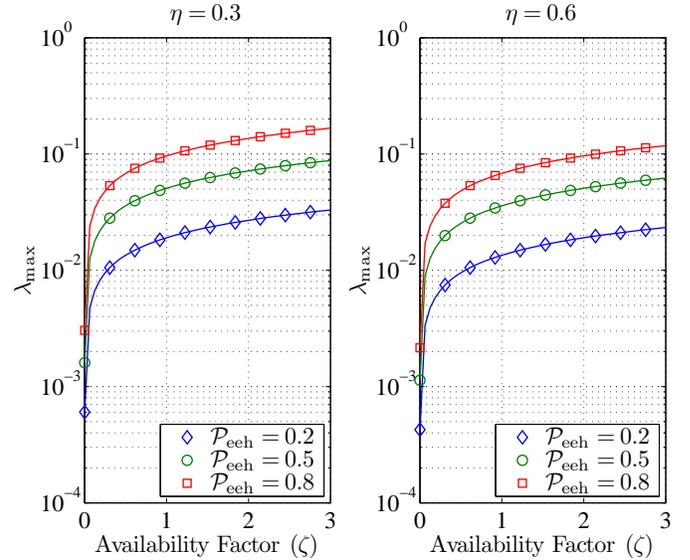}
\par\end{centering}

\caption{Maximal BS-deployment density over availability factor, with \label{fig:Maximal-BS-deployment-density}$\epsilon=0.3$
, $\rho=0.1$ and $p_{m}=0.02W$}

\end{figure}

We dipict curves of $\lambda_{{\rm max}}$ over $\zeta$ with constant
$\mathcal{P}_{{\rm eeh}}$s as figure \ref{fig:Maximal-BS-deployment-density}.
The left part sets the converter efficieny as $0.3$ and the left
part as $0.6$. The goal of this figure is to show how densely the
small cell BSs should be deployed to meet the requirement of avarage
EEH probability and availability factor. It is easy to find that $\lambda_{{\rm max}}$
is increased with the objective $\mathcal{P}_{{\rm eeh}}$ and the
availability factor. It can be observed that to achieve the objective
of $\mathcal{P}_{{\rm eeh}}=0.8$ and $\zeta$=1 (corrosponding to
level 2) with $\eta=0.3$, the needed BS density will be as much as
$10^{-1}$. Obviously it is too dense compared with current LTE-A
standard. Even if the converter efficiency $\eta$ is improved
to $0.6$, the density requirement is still impractical.

In conclusion, in lihght of above analysis the simultaneous information
and energy transfer for small cell LTE-A networks can only provide
very limited energy for the mobile terminals. That dose not mean the
SIET is infeasible, but the energy harvested from the BSs can charge
the secondary battery to prolong the lasting-time of terminals. Besides,
only the BS-deployment density can reach $10^{-1}$ or higher can
the harvested energy from BSs charge the primary battery or directly
power the mobile terminal.

\section{Conclusion\label{sec:Conclusion}}

In this paper we studied the feasibility of simultaneous information
and energy transfer in homogeneous small cell LTE-A networks. By stochastic
goemetry tools, we formulated the feasibility as maximaztion the available
average efficient-energy-harvesting(EEH) probability problem onditioned
on constraints of coverage probability , deployment density and transmit
power of the BSs. For tractability, we simplified the problem to a
special case with path loss exponent $\alpha=4$ and noise variance
$\sigma^{2}=0$. The solution for the special case shows that the
average EEH probability is increased with larger BS-deployment density
if other parameters like converter eficieny, energy harvesting threshold,
power splitting factor and user activity level is given . The numerical
results reveals that under current BS-deployment of LTE-A standard,
harvesting energy from BSs to charge the secondary battery of a hybrid-battery
powered terminal is feasible; but to sustain the basic system with
or without other computation tasks is infeasible.

In this paper, only the single-tier small cell networks is considered,
we will study the feasibility of SIET in the multi-tier small cell
networks in the future. The tradeoff between the harvested energy
and the inteference is also an interesting problem and needs further
study.


\begin{thebibliography}{10}
\providecommand{\url}[1]{#1}
\csname url@samestyle\endcsname
\providecommand{\newblock}{\relax}
\providecommand{\bibinfo}[2]{#2}
\providecommand{\BIBentrySTDinterwordspacing}{\spaceskip=0pt\relax}
\providecommand{\BIBentryALTinterwordstretchfactor}{4}
\providecommand{\BIBentryALTinterwordspacing}{\spaceskip=\fontdimen2\font plus
\BIBentryALTinterwordstretchfactor\fontdimen3\font minus
  \fontdimen4\font\relax}
\providecommand{\BIBforeignlanguage}[2]{{%
\expandafter\ifx\csname l@#1\endcsname\relax
\typeout{** WARNING: IEEEtran.bst: No hyphenation pattern has been}%
\typeout{** loaded for the language `#1'. Using the pattern for}%
\typeout{** the default language instead.}%
\else
\language=\csname l@#1\endcsname
\fi
#2}}
\providecommand{\BIBdecl}{\relax}
\BIBdecl

\bibitem{SUD11}
S.~Sudevalayam and P.~Kulkarni, ``Energy harvesting sensor nodes: Survey and
  implications,'' \emph{Communications Surveys Tutorials, IEEE}, vol.~13,
  no.~3, pp. 443--461, 2011.

\bibitem{VAR08}
L.~Varshney, ``Transporting information and energy simultaneously,'' in
  \emph{Information Theory, 2008. ISIT 2008. IEEE International Symposium on},
  2008, pp. 1612--1616.

\bibitem{KAW13}
S.~Kawasaki, ``The green energy harvesting winds by the rf/microwave power
  transmission,'' in \emph{Wireless Power Transfer (WPT), 2013 IEEE}, 2013, pp.
  111--114.

\bibitem{LIU13}
L.~Liu, R.~Zhang, and K.-C. Chua, ``Wireless information transfer with
  opportunistic energy harvesting,'' \emph{Wireless Communications, IEEE
  Transactions on}, vol.~12, no.~1, pp. 288--300, 2013.

\bibitem{SIX13}
S.~Yin, E.~Zhang, J.~Li, L.~Yin, and S.~Li, ``Throughput optimization for
  self-powered wireless communications with variable energy harvesting rate,''
  in \emph{Wireless Communications and Networking Conference (WCNC), 2013
  IEEE}, 2013, pp. 830--835.

\bibitem{ZHANG13}
R.~Zhang and C.~K. Ho, ``Mimo broadcasting for simultaneous wireless
  information and power transfer,'' \emph{Wireless Communications, IEEE
  Transactions on}, vol.~12, no.~5, pp. 1989--2001, 2013.

\bibitem{huang12}
K.~Huang and V.~K. Lau, ``Enabling wireless power transfer in cellular
  networks: architecture, modeling and deployment,'' \emph{arXiv preprint
  arXiv:1207.5640}, 2012.

\bibitem{Dol10}
A.~Dolgov, R.~Zane, and Z.~Popovic, ``Power management system for online low
  power rf energy harvesting optimization,'' \emph{Circuits and Systems I:
  Regular Papers, IEEE Transactions on}, vol.~57, no.~7, pp. 1802--1811, 2010.

\bibitem{Naka13}
T.~Nakamura, S.~Nagata, A.~Benjebbour, Y.~Kishiyama, T.~Hai, S.~Xiaodong,
  Y.~Ning, and L.~Nan, ``Trends in small cell enhancements in lte advanced,''
  \emph{Communications Magazine, IEEE}, vol.~51, no.~2, pp. 98--105, 2013.

\bibitem{Muk11}
S.~Mukherjee, ``Ue coverage in lte macro network with mixed csg and open access
  femto overlay,'' in \emph{Communications Workshops (ICC), 2011 IEEE
  International Conference on}, 2011, pp. 1--6.

\bibitem{martin12}
M.~Haenggi, \emph{Stochastic geometry for wireless networks}.\hskip 1em plus
  0.5em minus 0.4em\relax Cambridge University Press, 2012.

\bibitem{Scor13}
S.~Scorcioni, L.~Larcher, A.~Bertacchini, L.~Vincetti, and M.~Maini, ``An
  integrated rf energy harvester for uhf wireless powering applications,'' in
  \emph{Wireless Power Transfer (WPT), 2013 IEEE}, 2013, pp. 92--95.

\bibitem{Balasub09}
\BIBentryALTinterwordspacing
N.~Balasubramanian, A.~Balasubramanian, and A.~Venkataramani, ``Energy
  consumption in mobile phones: a measurement study and implications for
  network applications,'' in \emph{Proceedings of the 9th ACM SIGCOMM
  conference on Internet measurement conference}, ser. IMC '09.\hskip 1em plus
  0.5em minus 0.4em\relax New York, NY, USA: ACM, 2009, pp. 280--293. [Online].
  Available: \url{http://doi.acm.org/10.1145/1644893.1644927}
\BIBentrySTDinterwordspacing

\bibitem{Carroll10}
\BIBentryALTinterwordspacing
A.~Carroll and G.~Heiser, ``An analysis of power consumption in a smartphone,''
  in \emph{Proceedings of the 2010 USENIX conference on USENIX annual technical
  conference}, ser. USENIXATC'10.\hskip 1em plus 0.5em minus 0.4em\relax
  Berkeley, CA, USA: USENIX Association, 2010, pp. 21--21. [Online]. Available:
  \url{http://dl.acm.org/citation.cfm?id=1855840.1855861}
\BIBentrySTDinterwordspacing

\end{thebibliography}

\end{document}